\documentclass[fleqn,twoside]{article}
\usepackage{espcrc2,graphicx}
\title{Multiloop calculations in HQET}

\author{A.G.~Grozin\address{Budker Institute of Nuclear Physics, Novosibirsk}}
       
\begin{document}

\begin{abstract}
Recently, algorithms for calculation of 3-loop propagator diagrams in HQET and
on-shell QCD with a heavy quark have been constructed and implemented.
These algorithms (based on integration by parts recurrence relations)
reduce an arbitrary diagram to a combination of a finite number of basis integrals.
Here I discuss various ways to calculate non-trivial bases integrals,
either exactly or as expansions in $\varepsilon$.
Some integrals of these two classes are related to each other by inversion,
which provides a useful cross-check.
\vspace{1pc}
\end{abstract}

\maketitle

I presented a review talk about multiloop calculations in HQET
at this conference in Pisa in 1995~\cite{BG:95}.
Methods of calculation of two-loop propagator diagrams in HQET~\cite{BG:91}
and on-shell massive QCD~\cite{B:92,FT:92},
based on integration by parts~\cite{CT:81},
were discussed there.
Recently, three-loop HQET~\cite{G:00} and on-shell~\cite{MR:00} algorithms
have been constructed.
Here I discuss this substantial progress.

\section{Three-loop massless diagrams}
\label{QCD}

First, I briefly remind you the classic method of calculation
of 3-loop massless propagator diagrams.
There are 3 generic topologies of such diagrams.
They can be reduces, using integration by parts, to 6 basis integrals~\cite{CT:81}.
This algorithm is implemented in the package Mincer~\cite{Mincer}
(first written in SCHOONSCHIP~\cite{Sch} and later rewritten in FORM~\cite{Form}),
and in the package Slicer~\cite{Sli} written in REDUCE~\cite{H:99,G:97}.

Four basis integrals are trivial.
One is a two-loop diagram with a non-integer power of the middle line.
It can be found as a particular case of a more general expression~\cite{K:96,BGK:97}
for the the two-loop diagram with three non-integer powers
via a hypergeometric ${}_3F_2$ function of the unit argument,
with indices tending to integers at $\varepsilon\to0$.
There is a rather straightforward algorithm for expanding such functions in $\varepsilon$,
with coefficients expressed via multiple $\zeta$-values.
I have implemented it in REDUCE in the summer of 2000,
some results produced by this program are published in~\cite{G:01}.
It is clearly presented as Algorithm A in~\cite{MUW:02};
this paper also contains other, more complicated, algorithms.
The algorithms of~\cite{MUW:02} are implemented in the C++ library
nestedsums~\cite{W:02} based on the computer-algebra library GiNaC~\cite{BFK:00}.
This implementation is very convenient;
unfortunately, it requires one to install an outdated version of GiNaC.
The Algorithm A seems to be also implemented in FORM~\cite{V:99},
but I could not understand how to use it.
Using my REDUCE procedure or nestedsums~\cite{W:02},
one can quickly find as many terms of expansion of this basis integral
in $\varepsilon$ as needed, in terms of multiple $\zeta$-values.
They can be expressed, up to weight 9, via a minimum set of independent
$\zeta$-values, using the results of~\cite{B:96,V:99}.
The two-loop diagram with a non-integer power of the middle line
can also be expressed~\cite{K:85} via an ${}_3F_2$ function of the argument $-1$.
Expanding this expression in $\varepsilon$ (say, using nestedsums~\cite{W:02}),
we encounter more general Euler--Zagier sums,
which were also considered in~\cite{B:96}.
Reducing them to the minimal basis, we obtain, of course,
the same $\varepsilon$-expansion of our basis integral.

Using this expansion and integration-by-parts relations,
it is easy to recover the well-known result for the 3-loop ladder diagram,
which is finite $\varepsilon=0$: $20\zeta(5)+\mathcal{O}(\varepsilon)$.
The last and most difficult basis diagram is non-planar.
It is also finite at $\varepsilon=0$.
Using gluing of its external vertices~\cite{CT:81},
one can easily understand that it has the same value $20\zeta(5)$
at $\varepsilon=0$.
There is no easy way to find further terms of its $\varepsilon$- expansion.

\begin{figure*}
\begin{picture}(92,50)
\put(46,25){\makebox(0,0){\includegraphics{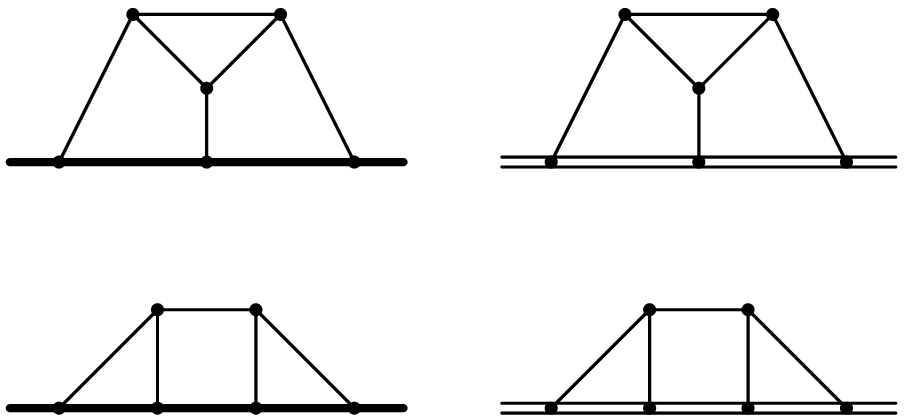}}}
\put(13.5,26){\makebox(0,0)[b]{$n_1$}}
\put(28.5,26){\makebox(0,0)[b]{$n_2$}}
\put(63.5,26){\makebox(0,0)[b]{$n_1$}}
\put(78.5,26){\makebox(0,0)[b]{$n_2$}}
\put(8.5,37.5){\makebox(0,0)[r]{$n_3$}}
\put(33,37.5){\makebox(0,0)[l]{$n_4$}}
\put(58.5,39){\makebox(0,0)[r]{$d-n_1-n_3$}}
\put(57,36){\makebox(0,0)[r]{${}-n_5-n_7$}}
\put(83,39){\makebox(0,0)[l]{$d-n_2-n_4$}}
\put(84,36){\makebox(0,0)[l]{${}-n_5-n_8$}}
\put(22,33.75){\makebox(0,0)[l]{$n_5$}}
\put(72,33.75){\makebox(0,0)[l]{$n_5$}}
\put(17.7,39){\makebox(0,0)[r]{$n_7$}}
\put(25,39){\makebox(0,0)[l]{$n_8$}}
\put(67.7,39){\makebox(0,0)[r]{$n_7$}}
\put(75,39){\makebox(0,0)[l]{$n_8$}}
\put(21,46){\makebox(0,0)[b]{$n_6$}}
\put(71,46){\makebox(0,0)[b]{$d-n_6-n_7-n_8$}}
\put(46,30){\makebox(0,0){$=$}}
\put(11,1){\makebox(0,0)[b]{$n_1$}}
\put(21,1){\makebox(0,0)[b]{$n_3$}}
\put(31,1){\makebox(0,0)[b]{$n_2$}}
\put(61,1){\makebox(0,0)[b]{$n_1$}}
\put(71,1){\makebox(0,0)[b]{$n_3$}}
\put(81,1){\makebox(0,0)[b]{$n_2$}}
\put(9.5,10){\makebox(0,0)[r]{$n_4$}}
\put(32,10){\makebox(0,0)[l]{$n_5$}}
\put(59.5,12){\makebox(0,0)[r]{$d-n_1-n_4$}}
\put(57.5,9){\makebox(0,0)[r]{${}-n_6$}}
\put(82,12){\makebox(0,0)[l]{$d-n_2-n_5$}}
\put(83,9){\makebox(0,0)[l]{${}-n_7$}}
\put(17,10){\makebox(0,0)[l]{$n_6$}}
\put(25,10){\makebox(0,0)[r]{$n_7$}}
\put(67,10){\makebox(0,0)[l]{$n_6$}}
\put(75,10){\makebox(0,0)[r]{$n_7$}}
\put(21,16){\makebox(0,0)[b]{$n_8$}}
\put(71,16){\makebox(0,0)[b]{$d-n_3-n_6-n_7-n_8$}}
\put(46,5){\makebox(0,0){$=$}}
\end{picture}
\caption{Inversion relations}
\label{Fig}
\end{figure*}

\section{Three-loop HQET diagrams}
\label{HQET}

There are 10 generic topologies of 3-loop HQET propagator diagrams.
They can be reduced, using integration by parts, to 8 basis integrals~\cite{G:00}.
This algorithm is implemented in the REDUCE package Grinder~\cite{G:00},
available at http://www-ttp.physik.uni-karlsruhe.de/Progdata/ttp00/ttp00-01/.
Five basis integrals are trivial.
Two can be expressed via ${}_3F_2$ hypergeometric functions
of the unit argument~\cite{BB:94,G:00}.
Their expansions in $\varepsilon$ can be obtained in the same way
as in the massless case, the results are presented in~\cite{G:01}.
The last and most difficult basis integral was found in~\cite{G:01}
up to the finite term in $\varepsilon$,
using direct integration in the coordinate space.
More terms of its $\varepsilon$-expansion were recently obtained in~\cite{CM:02}
using inversion, as explained in the next Section.

\section{Three-loop on-shell diagrams}
\label{OS}

Calculations of on-shell diagrams with massive quarks in QCD
are necessary for obtaining coefficients in the HQET Lagrangian
and $1/m$ HQET expansions of QCD operators by matching.
There are 2 generic topologies of 2-loop on-shell propagator diagrams
with a single non-zero mass.
They can be reduced, using integration by parts, to 3 basis integrals.
This algorithm is implemented in the REDUCE package RECURSOR~\cite{B:92}
and the FORM package SHELL2~\cite{FT:92}.
Two basis integrals are trivial, and the third one is expressed
via two ${}_3F_2$ hypergeometric functions of the unit argument.
However, some of their indices tend to half-integers at $\varepsilon\to0$,
and the algorithm of expansion in $\varepsilon$ discussed in Sect.~\ref{QCD}
is not applicable.
This approach was used for QCD/HQET matching of heavy-light
quark currents~\cite{BG:95a} and chromomagnetic interaction~\cite{CG:97}.

The case when there is another non-zero mass
was systematically studied in~\cite{DG:99}.
There are 4 basis integrals, 2 of them trivial,
and 2 are expressed via ${}_3F_2$ hypergeometric functions
of the mass ratio squared.
Finite parts at $\varepsilon\to0$ are expressed via dilogarithms.
More terms of expansions of the general results~\cite{DG:99}
in $\varepsilon$ were recently obtained~\cite{AMR:02}.
The REDUCE package~\cite{DG:99} is available at
http://wwwthep.physik.uni-mainz.de/Publications/progdata/mzth9838/ Mm.red.

There are 11 generic topologies of 3-loop on-shell propagator diagrams
with a single non-zero mass (10 of them are the same as in HQET,
and one involves a heavy-quark loop).
They can be reduced, using integration by parts, to 18 basis integrals~\cite{MR:00}.
This algorithm is implemented as the FORM package SHELL3~\cite{MR:00}.
The basis integrals are mostly known from QED~\cite{LR:96}.

Some on-shell diagrams are related to HQET ones by inversion
of Euclidean integration momenta.
One- and two-loop relations were presented in~\cite{BG:95}.
Three-loop relations are shown in Fig.~\ref{Fig}.
The second of them was used in~\cite{CM:02}
to relate the the convergent ladder HQET diagram at $\varepsilon=0$
to the known on-shell ladder diagram.

\end{document}